\documentclass[aps,prl,amssymb,twocolumn,amsmath,superscriptaddress,showpacs,10pt]{revtex4}

\usepackage{graphicx}
\usepackage{dcolumn}
\usepackage{bm}
\usepackage{color}
\usepackage{epstopdf}

\begin{document}

\newcommand{\beq}{\begin{equation}}
\newcommand{\eeq}{\end{equation}}
\newcommand{\barr}{\begin{eqnarray}}
\newcommand{\earr}{\end{eqnarray}}

\newcommand{\REV}[1]{\textbf{\color{blue}[[#1]]}}
\newcommand{\GREEN}[1]{\textbf{\color{green}#1}}
\newcommand{\RED}[1]{\textrm{\color{red}#1}}
\newcommand{\rev}[1]{{\color{blue}#1}}

\newcommand{\andy}[1]{ }
\newcommand{\bmsub}[1]{\mbox{\boldmath\scriptsize $#1$}}

\def\R{\mathbb{R}}

\def\bra#1{\langle #1 |}
\def\ket#1{| #1 \rangle}
\def\sinc{\mathop{\text{sinc}}\nolimits}
\def\cV{\mathcal{V}}
\def\cH{\mathcal{H}}
\def\cT{\mathcal{T}}
\def\cM{\mathcal{M}}
\def\cN{\mathcal{N}}
\def\CW{\mathcal{W}}
\def\e{\mathrm{e}}
\def\ii{\mathrm{i}}
\def\d{\mathrm{d}}
\renewcommand{\Re}{\mathop{\text{Re}}\nolimits}
\newcommand{\tr}{\mathop{\text{Tr}}\nolimits}

\title{Greenberger-Horne-Zeilinger states and few-body Hamiltonians}

\author{Paolo Facchi}
\affiliation{Dipartimento di Matematica and MECENAS, Universit\`a di Bari, I-70125 Bari, Italy}
\affiliation{INFN, Sezione di Bari, I-70126 Bari, Italy}

\author{Giuseppe Florio}
\affiliation{Dipartimento di Fisica and MECENAS, Universit\`a di Bari, I-70126 Bari, Italy}
\affiliation{INFN, Sezione di Bari, I-70126 Bari, Italy}

\author{Saverio Pascazio}
\affiliation{Dipartimento di Fisica and MECENAS, Universit\`a di Bari, I-70126 Bari, Italy}
\affiliation{INFN, Sezione di Bari, I-70126 Bari, Italy}

\author{Francesco V. Pepe}
\affiliation{Dipartimento di Fisica and MECENAS, Universit\`a di Bari, I-70126 Bari, Italy}
\affiliation{INFN, Sezione di Bari, I-70126 Bari, Italy}

%\date{\today}

\begin{abstract}
The generation of Greenberger-Horne-Zeilinger (GHZ) states is a
crucial problem in quantum information.
We derive general conditions for obtaining GHZ states as
eigenstates of a Hamiltonian. In general, degeneracy cannot be
avoided if the Hamiltonian contains $m$-body interaction terms
with $m \leq 2$ and a number of qubits strictly larger than 4. As
an application, we explicitly construct a two-body 4-qubit
Hamiltonian and a three-body 5-qubit Hamiltonian that exhibit a
GHZ as a nondegenerate eigenstate.
\end{abstract}


\maketitle

The use of quantum mechanics for improving tasks such as
communication, computation and cryptography \cite{nielsen} is
based on the availability of highly  entangled states 
\cite{h4,entanglement,entanglementrev,adessorev}. It is therefore of primary
importance to obtain reliable strategies for their generation.
Among others, GHZ states \cite{ghz} represent a paradigmatic
example of multipartite entangled states. In particular, in the
case of three qubits, these states contain purely tripartite
entanglement \cite{cirac} and do not retain any bipartite
entanglement when one of the qubits is traced out, thus maximizing
the residual tangle \cite{multipart1}.

The experimental realization of GHZ states
\cite{esperimentighz,esperimentighz2,esperimentighz3,esperimentighz4},
most recently with 14 qubits \cite{14} has paved the way towards
realistic implementation of quantum protocols. In these experiments
a bottom-up approach is employed, whereby individual quantum systems
(trapped particles, photons, cavities) are combined and
manipulated.
As the number of controllable qubits increases,
the generation of GHZ states require the use of quantum
operations, whose feasibility strongly depends on the physical
system used (optical, semiconductor or superconductor based
\cite{molmer,generationghz}). In the case of the recent trapped-ion implementation
\cite{14}, the problem is additionally complicated
by the presence of correlated Gaussian phase noise, that provokes
``superdecoherence", by which decay scales quadratically with the
number of qubits. It becomes therefore necessary to manipulate and
control state fidelity and dynamics over sufficiently long
timescales.

In principle, an alternative scheme for the implementation of GHZ
states would consist in its encoding into one of the eigenstates
(possibly the fundamental one) of a suitable Hamiltonian.
For instance, in \cite{buzek} it
was shown that for the quantum Ising model in a transverse field
the ground state is approximately a GHZ state if the strength of
the field goes to infinity. Moreover, a proper choice of local
fields for an Heisenberg-like spin model can yield a ground state
which is, again, approximately GHZ \cite{loss1,loss2}.

On the other hand, it would be interesting  to understand what are
the requirements to obtain an \emph{exact} GHZ state as an
eigenstate of a quantum Hamiltonian. In this Letter we will
address this problem and find rigorous conditions for the encoding
of GHZ states into one of the eigenstates of a Hamiltonian that contains few-body coupling
terms.

Let
\begin{equation}
\label{eq:ghz}
|G_\pm^{n}\rangle=\frac{1}{\sqrt{2}} \left(|0\rangle^{\otimes
n}\pm |1\rangle^{\otimes n}\right)
\end{equation}
be GHZ states, where $\sigma^z|i\rangle=(-1)^i | i \rangle$
defines the computational basis, with $i=0,1$ and $\sigma^z$ the
third Pauli matrix. As a preliminary remark, we notice that it is
trivial to find Hamiltonians involving $n$-body interaction terms,
whose nondegenerate ground state is $|G_+^{n}\rangle$: the
simplest example is $E_0|G_+^{n}\rangle\langle G_+^{n}|$, with
$E_0<0$. On the other hand, we can ask whether it is possible for
$|G_+^{n}\rangle$ to be the nondegenerate ground state, even if
the Hamiltonian involves at most $m$-body interaction terms (with
$m<n$). One can easily see that this is not possible. The reason
lies in the fact that $|G_+^{n}\rangle$ and $|G_-^{n}\rangle$
share the same $m$-body reduced density matrices, and thus the
same expectation values on $m$-body interaction terms. If
$|G_+^{n}\rangle$ is a ground state, also $|G_-^{n}\rangle$ must
be a ground state. This is a special case of a result proved in
\cite{nielsen2}.

Thus, we relax our initial requirement and try to understand
whether $|G_+^{n}\rangle$ can be a nondegenerate excited
eigenstate for some $m$-body Hamiltonian. More specifically, we
search for a limiting value  $m_n^*$, depending on the number $n$
of qubits in the system, such that, if the Hamiltonian involves
$m$-body interaction terms (with $m<m_n^*$), $|G_+^{n}\rangle$
cannot be a nondegenerate eigenstate, otherwise the task becomes
possible. The most generic $m$-body Hamiltonian acting on the
Hilbert space of $n$ qubits can be written as
\begin{equation}\label{eq:hm}
H^{(m)}=\sum_{j_1=1}^n\ldots\sum_{j_m=1}^n\sum_{\alpha_1}
\ldots\sum_{\alpha_m} J_{j_1\ldots
j_m}^{\alpha_1\ldots\alpha_m}
\sigma_{j_1}^{\alpha_1}\ldots\sigma_{j_m}^{\alpha_m}
\end{equation}
with $\alpha_i=0,x,y,z$, $\sigma_i^0\equiv\openone_i$ being the
identity operator, $\sigma_i^\alpha$ the Pauli matrices acting on
the Hilbert space of qubit $i$ and $J$'s real numbers. Terms
involving only identities and an even number of $\sigma^z$'s map
$|G_+^{n}\rangle$ on the subspace spanned by itself. On the other
hand, terms involving other (products of) Pauli matrices map
$|G_+^{n}\rangle$ onto an orthogonal subspace. The action of
$H^{(m)}$ on $\ket{G_+^n}$ is
\begin{equation}\label{eq:hmg+}
H^{(m)}|G_+^{n}\rangle=\epsilon|G_+^{n}\rangle+|\Psi^{(m)}\rangle,
\end{equation}
where $\epsilon$ is a multiplicative constant and
$|\Psi^{(m)}\rangle$ is an unnormalized state vector satisfying
\begin{equation}\label{eq:ort+}
\langle\Psi^{(m)}|G_+^{n}\rangle=0 .
\end{equation}
Since the action of the Hamiltonian (\ref{eq:hm}) consists in
inverting spins and changing the relative sign of $\ket{G_+^n}$,
the vector $|\Psi^{(m)}\rangle$ can be expressed in a convenient
way by introducing a new notation. Let
\begin{equation}
\mathcal{N}=\left(1,2,\ldots,n\right)
\end{equation}
be the ordered set of naturals from $1$ to $n$, and let
\begin{equation}\label{eq:indices}
\mathcal{I}=\left( i_1, i_2,\ldots,i_l\right)
\end{equation}
denote a multi-index, whose elements range from $1$ to $n$ and
satisfy $i_1<i_2<\ldots<i_l$. The cardinality $|\mathcal{I}|=l$
verifies
\begin{equation}\label{eq:indlength}
1\leq |\mathcal{I}|\leq m<n.
\end{equation}
We now define a set of normalized state vectors, depending on the
choice of the multi-index $\mathcal{I}$ and on the sign
$\sigma=\pm$: \barr\label{eq:gtilde}
|\tilde{G}_{\sigma,\mathcal{I}}^n\rangle&=&\frac{1}{\sqrt{2}}\left[
\left(\bigotimes_{i\in\mathcal{I}}|1\rangle_i\bigotimes_{j\in\mathcal{N}/\mathcal{I}}|0\rangle_j\right)\right.\nonumber\\
&+&\sigma\left.\left(
\bigotimes_{i\in\mathcal{I}}|0\rangle_i\bigotimes_{j\in\mathcal{N}/\mathcal{I}}|1\rangle_j
\right)\right] \earr The state
$|\tilde{G}_{\sigma,\mathcal{I}}^n\rangle$ differs from
$|G_+^n\rangle$ in that spins corresponding to the indices in
$\mathcal{I}$ are reversed in both computational basis vectors in
the superposition $\ket{G_+^n}$. This means that
$|\tilde{G}_{\sigma,\mathcal{I}}^n\rangle=\ket{G_+^n}$ if
$\mathcal{I}$ is the empty set. Moreover, the relative phase of
the two vectors can be positive or negative, according to the sign
$\sigma$. Thus, the vector $|\Psi^{(m)}\rangle$ in Eq.\
(\ref{eq:hmg+}) can be expressed as
\begin{equation}\label{eq:psim}
|\Psi^{(m)}\rangle=b_0|G_-^n\rangle+\sum_{\mathcal{I}} \left(
a_{\mathcal{I}}|\tilde{G}_{+,\mathcal{I}}^n\rangle +
b_{\mathcal{I}} |\tilde{G}_{-,\mathcal{I}}^n\rangle \right).
\end{equation}
The coefficients $a_{\mathcal{I}}$, $b_{\mathcal{I}}$ and $b_0$
are functions of the parameters of the Hamiltonian
(\ref{eq:hm}). It is obvious that, if they can all be set to zero
by a proper choice of $H^{(m)}$,
$|G_+^{(m)}\rangle$ will be an eigenstate of the Hamiltonian. A problem
arises, however, if we take into account the antisymmetric
state $|G_-^{(m)}\rangle$. The action of $H^{(m)}$ on
this vector reads
\begin{equation}\label{eq:hmg-}
H^{(m)}|G_-^{n}\rangle=\epsilon|G_-^{n}\rangle+|\Phi^{(m)}\rangle,
\end{equation}
where $|\Phi^{(m)}\rangle$ is orthogonal to $|G_-^{n}\rangle$ and
can be decomposed as
\begin{equation}\label{eq:phim}
|\Phi^{(m)}\rangle=b_0|G_+^n\rangle+\sum_{\mathcal{I}} \left(
a_{\mathcal{I}}|\tilde{G}_{-,\mathcal{I}}^n\rangle +
b_{\mathcal{I}} |\tilde{G}_{+,\mathcal{I}}^n\rangle \right).
\end{equation}
If all the coefficients in Eq.\ (\ref{eq:psim}) are set to zero,
this will result in the cancellation of $|\Phi^{(m)}\rangle$. As a
consequence, $|G_+^n\rangle$ and $|G_-^n\rangle$ will be
degenerate eigenstates (with eigenvalue $\epsilon$). Thus, if the
sufficient conditions
\begin{eqnarray}
\label{eq:cond1}
& & b_0=0, \\
\label{eq:cond2}
& & a_{\mathcal{I}}=0\,,\quad b_{\mathcal{I}}=0
\end{eqnarray}
are also \textit{necessary} for $|G_+^n\rangle$ to be an
eigenstate of $H^{(m)}$, degeneracy is unavoidable. We notice
that, since the
following equality holds
\begin{equation}
\langle G_-^n | \tilde{G}_{\sigma,\mathcal{I}}^n \rangle=0\quad \forall\mathcal{I} \quad \mbox{and} \quad \forall\sigma,
\end{equation}
Eq.\ (\ref{eq:cond1}) is always a necessary condition.

Let us start considering the case in which the Hamiltonian
(\ref{eq:hm}) contains interaction terms up to $m$-body such that
\beq m<m_n^*\equiv [(n+1)/2] \eeq with $[\cdot]$ denoting the
integer part. Following Eq.\ (\ref{eq:indlength}), the sum in the
decomposition of $|\Psi^{(m)}\rangle$ and $|\Phi^{(m)}\rangle$
runs over all the multi-indices whose length satisfies
\begin{equation}\label{eq:indlengthA}
1\leq |\mathcal{I}|\leq m<m_n^*.
\end{equation}
If this inequality holds, the following orthogonality
relations are verified:
\begin{equation}\label{eq:orthog}
\langle \tilde{G}_{\sigma_1,\mathcal{I}_1}^n |
\tilde{G}_{\sigma_2,\mathcal{I}_2}^n \rangle=0 \quad \text{if }
\mathcal{I}_1\neq \mathcal{I}_2 \text{ or } \sigma_1\neq\sigma_2 .
\end{equation}
Thus, Eq.\ (\ref{eq:cond2}) is a necessary condition to cancel
$|\Psi^{(m)}\rangle$ and make $|G_+^n\rangle$ an eigenstate of
$H^{(m)}$. In this case, however, $|G_+^n\rangle$ and
$|G_-^n\rangle$ are eigenstates corresponding to the same
eigenvalue. We can conclude that, if the Hamiltonian of a qubit
system involves terms coupling less than $m_n^*= [(n+1)/2]$ spins,
the GHZ state $\ket{G_+^n}$, and any equivalent state by local
unitaries, cannot be a nondegenerate eigenstate. If $\ket{G_+^n}$
is an eigenstate for some Hamiltonian $H^{(m)}$, it must be at
least two-fold degenerate.

On the other hand, if $m=m_n^*$ degeneracy can be avoided.
Actually, in this case some conditions in Eq. (\ref{eq:cond2}) are
no longer necessary and, therefore, the orthogonality relations in
Eq.\ (\ref{eq:orthog}) hold if inequality (\ref{eq:indlengthA}) is
satisfied. However, a new relation emerges connecting
$|\tilde{G}_{\sigma,\mathcal{I}}^n\rangle$ states corresponding to
multi-indices of length $m_n^*$ and $(n-m_n^*)$ (which is equal to
$m_n^*$ for even $n$ and to $m_n^*-1$ for odd $n$). Indeed,
reversing $m_n^*$ spins in $\ket{G_+^n}$ is completely equivalent
to reversing the other $n-m_n^*$ ones. Instead, if the same
operations are applied on the antisymmetric state $\ket{G_-^n}$,
they will differ only by an overall sign. Thus, we have the
following relations
\begin{equation}\label{eq:gtildelim}
|\tilde{G}_{\pm,\mathcal{I}}^n\rangle=\pm|\tilde{G}_{\pm,\mathcal{N}/\mathcal{I}}^n\rangle\quad
\text{ if } |\mathcal{I}|=m_n^*,n-m_n^*.
\end{equation}
While conditions (\ref{eq:cond2}) still hold for
$|\mathcal{I}|<\min(m_n^*,n-m_n^*)$, for larger values of
$|\mathcal{I}|$ one should use
\begin{equation}\label{eq:cond2b}
\left\{
\begin{array}{l}
a_{\mathcal{I}}=-a_{\mathcal{N}/\mathcal{I}} \\
b_{\mathcal{I}}=b_{\mathcal{N}/\mathcal{I}}
\end{array}
\right. \qquad \mbox{if}\quad |\mathcal{I}|=m_n^*,n-m_n^*.
\end{equation}
Thus, in order to cancel $|\Psi^{(m)}\rangle$, it is no longer
necessary to set all the coefficient $a_{\mathcal{I}}=0$ and
$b_{\mathcal{I}}=0$ in Eq.\ (\ref{eq:psim}) because this would give
a degeneracy (remember that, by the same conditions, one would have
$|\Phi^{(m)}\rangle=0$). Instead, by using Eq.(\ref{eq:cond2b}), the
vector $|\Phi^{(m)}\rangle$ in Eq.(\ref{eq:phim}) becomes
\begin{equation}\label{eq:phimbar}
|\bar{\Phi}^{(m)}\rangle\equiv\sum_{|\mathcal{I}|=m_n^*,n-m_n^*}
\left( a_{\mathcal{I}}|\tilde{G}_{-,\mathcal{I}}^n\rangle +
b_{\mathcal{I}} |\tilde{G}_{+,\mathcal{I}}^n\rangle \right),
\end{equation}
which is generally different from the null vector. If, for some
values of the parameters in the Hamiltonian (\ref{eq:hm}), the
conditions (\ref{eq:cond2b}) are satisfied without cancelling
$|\bar{\Phi}^{(m)}\rangle$, the GHZ state $|G_+^n\rangle$ can, at
least in principle, be a nondegenerate eigenstate of an
Hamiltonian with interaction terms coupling no more than
$m_n^*=[(n+1)/2]$ qubits. As a further remark, we notice that this
result does not ensure that $|G_+^n\rangle$ is nondegenerate. The
absence of degeneracy can be excluded only by the explicit
solution of the Hamiltonian.

The case $m_n^*<m<n$ is analogous to the previous one, since
conditions of the type (\ref{eq:gtildelim}) hold for all
multi-indices $\mathcal{I}$ that satisfy $n-m\leq
|\mathcal{I}|\leq m$. Following the same procedure as in the case
$m=m_n^*$, we find that the degeneracy of the eigenspace spanned
by $|G_+^n\rangle$ and $|G_-^n\rangle$ can still be avoided.

We will now consider an interesting application of the previous
results. In the following example, we will focus our attention on
the symmetric GHZ state $|G_+^n\rangle$ and the case
\barr\label{eq:four} n=4, \quad m=m_4^*=2 \earr and will show that
state
$|G_+^4\rangle=\left(|0000\rangle+|1111\rangle\right)/\sqrt{2}$
can be a nondegenerate eigenstate of a two-body Hamiltonian. We
restrict our attention to Hamiltonians involving only two body
coupling along the $x$ and $z$ axes, and we add the condition of
nearest-neighbour couplings on a ring:
\begin{equation}\label{eq:h4}
H^{(2)}=\sum_{i=1}^4 \left( J_{i}^x \sigma_i^x\sigma_{i+1}^x +
J_{i}^z \sigma_i^z\sigma_{i+1}^z \right) .
\end{equation}
In Eq.\ (\ref{eq:h4}) we have used periodic boundary conditions
$\bm{\sigma}_5\equiv\bm{\sigma_1}$. We notice that interaction
terms of the form $\sigma_i^z\sigma_{i+1}^z$ leave $|G_+^4\rangle$
invariant, while the terms $\sigma_i^x\sigma_{i+1}^x$ reverse two
nearest-neighbour spins. Since, as in Eq.\ (\ref{eq:gtildelim}),
we have
\begin{equation}\label{eq:gtildelim4}
|\tilde{G}_{\pm,(1,2)}^4\rangle=\pm|\tilde{G}_{\pm,(3,4)}^4\rangle, \;\; |\tilde{G}_{\pm,(2,3)}^4\rangle=\pm|\tilde{G}_{\pm,(1,4)}^4\rangle,
\end{equation}
the action of $H^{(2)}$ on the four-qubits GHZ state
$|G_+^4\rangle$ reads
\begin{equation}\label{h2g4}
H^{(2)}|G_+^4\rangle=\left(\sum_{i=1}^4
J_i^z\right)|G_+^4\rangle+\sum_{i=1}^2\left(J_i^x+J_{i+2}^x\right)|\tilde{G}_{\pm,(i,i+1)}^4\rangle.
\end{equation}
Thus $|G_+^4\rangle$ is an eigenstate of $H^{(2)}$ if and only if
\begin{equation}
\left\{
\begin{array}{l}
J_1^x=-J_3^x \\ J_2^x=-J_4^x .
\end{array}
\right.
\end{equation}
Under these conditions, the Hamiltonian acts on the antisymmetric
combination
$|G_-^4\rangle=\left(|0000\rangle-|1111\rangle\right)/\sqrt{2}$ as
\begin{equation}
H^{(2)}|G_-^4\rangle=\left(\sum_{i=1}^4
J_i^z\right)|G_-^4\rangle+2\sum_{i=1}^2 J_i^x
|\tilde{G}_{\pm,(i,i+1)}^4\rangle.
\end{equation}
If $J_1^x\ne 0$ or $J_2^x\ne 0$, $|G_-^4\rangle$ is not an
eigenstate. We explicitly solve a simple model, with $J_i^z\equiv
J^z/4$ for all $i$ and $J_1^x=J_2^x\equiv J^x/4(=-J_3^x=-J_4^x)$.
If the coupling constants are not zero and $J^x\neq J^z$,
$|G_+^4\rangle$ is a nondegenerate eigenstate, corresponding to
the eigenvalue $J^z$. It is remarkable, however, that this model
has three other eigenstates which are equivalent to
$|G_+^4\rangle$ by local unitaries, corresponding to the
eigenvalues $-J^z$ and $\pm J^x$.  As expected, none of them can
be the nondegenerate ground state, since the ground-state energy
is $\epsilon_0=\sqrt{(J^x)^2+(J^z)^2}$. $\ket{G_+^4}$ is the
(nondegenerate) first excited state if $J^z< 0$ and
$-1<J^x/J^z<1$. Incidentally, for different ranges of the
parameters the first excited state of this Hamiltonian is one of
the three eigenstates which are locally equivalent to
$\ket{G_+^4}$.

The five-qubit GHZ state
$\ket{G_+^5}=(\ket{00000}+\ket{11111})/\sqrt{2}$ needs at least
three-body interactions to be the nondegenerate eigenstate of any
Hamiltonian ($m_5^*=3$). It is straightforward to check that
$\ket{G_+^5}$ is an eigenstate of
\begin{equation}\label{eq:h5}
H^{(3)}=\frac{J^z}{5} \sum_{i=1}^5 \sigma_i^z\sigma_{i+1}^z +
\frac{J^x}{5}\sum_{i=1}^5 \left(
\sigma_i^x\sigma_{i+1}^x\sigma_{i+2}^x-\sigma_i^x\sigma_{i+1}^x
\right)
\end{equation}
with eigenvalue $J^z$ (periodic boundary conditions are assumed).
The conditions for $\ket{G_+^5}$ to be a nondegenerate eigenstate
are easily worked out by diagonalizing $H^{(3)}$. It is the
nondegenerate first excited eigenstate if $J^z<0$, $J^x\neq 0$ and
\begin{equation}
-2+\frac{2}{\sqrt{3}}<\frac{J^x}{J^z}<\frac{1}{6}\left[
\sqrt{2(75+7\sqrt{5})}-(7+\sqrt{5}) \right].
\end{equation}
$\ket{G_+^5}$ can be the ground state only if $J^x=0$, but in this
case it is twice degenerate.

In conclusion, we investigated general conditions such that GHZ
states (\ref{eq:ghz}) are nondegenerate in the spectrum of a
Hamiltonian. We showed that if the Hamiltonian acting on the
Hilbert space of $n$ qubits involves terms that couple at most $m$
qubits, it is impossible to have a nondegenerate GHZ eigenstate if
$m<m_n^*$ with $m_n^*=\left[ (n+1)/2\right]$. If $m\ge m_n^*$,
degeneracy can in principle be absent.

The difficulty in obtaining GHZ states as ground states (or even
eigenstates) of Hamiltonians that involve only few-body
interactions is in accord with previous results \cite{pepe} and
seems to be a characteristic trait of multipartite entanglement.
It would be interesting, also in view of applications, to
investigate the existence of general conditions for obtaining
approximate GHZ states for an arbitrary number of qubits by making
use of few-body Hamiltonians.

\acknowledgments
P.F.\ and G.F.\ acknowledge support through the project IDEA of University of Bari.

%%%%%%%%%%%%%%%%%%%%%%%%%%%%%%%%%%%%


\begin{thebibliography}{99}

\bibitem{nielsen}
 M.A. Nielsen and I.L. Chuang, {\it Quantum Computation and Quantum Information} (Cambridge University Press, Cambridge, 2000).

\bibitem{entanglement}
W. K. Wootters, Quantum Inf. Comp. \textbf{1}, 27 (2001).

\bibitem{entanglementrev}
L. Amico, R. Fazio, A. Osterloh and V. Vedral, Rev. Mod. Phys.
\textbf{80}, 517 (2008).

\bibitem{h4}
R. Horodecki, P. Horodecki, M. Horodecki and K. Horodecki, Rev.
Mod. Phys. \textbf{81}, 865 (2009).

\bibitem{adessorev}
G. Adesso and F. Illuminati,
J. Phys. A: Math. Theor. \textbf{40}, 7821 (2007).

\bibitem{ghz}
D. M. Greenberger, M. Horne, and A. Zeilinger, Am. J. Phys. \textbf{58}, 1131 (1990).

\bibitem{cirac}
W. Dur, G. Vidal, and J. I. Cirac, Phys. Rev. A \textbf{62}, 062314 (2000).

\bibitem{multipart1}
V. Coffman, J. Kundu and W. K. Wootters, Phys. Rev. A
\textbf{61}, 052306 (2000).

\bibitem{esperimentighz}
D. Bouwmeester, J.W. Pan, M. Daniell, H. Weinfurter, and A. Zeilinger, Phys. Rev. Lett. \textbf{82}, 1345 (1999).

\bibitem{esperimentighz2}
D. Leibfried, E. Knill, S. Seidelin, J. Britton, R. B. Blakestad, J. Chiaverini, D. B. Hume, W. M. Itano, J. D. Jost, C. Langer, R. Ozeri, R. Reichle and D. J. Wineland, Nature \textbf{438}, 639 (2005).

\bibitem{esperimentighz3}
W.-B. Gao, C.-Y. Lu, X.-C. Yao, P. Xu, O. Guhne, A. Goebel, Y.-A. Chen, C.-Z. Peng, Z.-B. Chen and J.-W. Pan, Nat. Phys. \textbf{6}, 331 (2010).

\bibitem{esperimentighz4}
R. Prevedel, G. Cronenberg, M. S. Tame, M. Paternostro, P. Walther, M. S. Kim, and A. Zeilinger, Phys. Rev. Lett. \textbf{103}, 020503 (2009).

\bibitem{14}
T. Monz, P. Schindler, J.T. Barreiro, M. Chwalla, D. Nigg, W.A. Coish, M. Harlander, W. H\"ansel, M. Hennrich and R. Blatt,
Phys. Rev. Lett. \textbf{106}, 130506 (2011).

\bibitem{molmer}
K. M\o lmer and A. S\o rensen, Phys. Rev. Lett. \textbf{82}, 1835 (1999).

\bibitem{generationghz}
L. F. Wei, Yu-xi Liu, and F. Nori, Phys. Rev. Lett. \textbf{96}, 246803 (2006); Ying-Dan Wang, S. Chesi, D. Loss, and C. Bruder
Phys. Rev. B \textbf{81}, 104524 (2010).

\bibitem{buzek}
P. Stelmachovic and V. Buzek, Phys. Rev. A \textbf{70}, 032313 (2004).

\bibitem{loss1}
B. Rothlisberger, J. Lehmann, D. S. Saraga, P. Traber,
and D. Loss, Phys. Rev. Lett. \textbf{100}, 100502 (2008).

\bibitem{loss2}
 B. Rothlisberger, J. Lehmann, and D. Loss, Phys. Rev. A \textbf{80}, 042301 (2009).

\bibitem{nielsen2}
H. L. Haselgrove, M. A. Nielsen and T. J. Osborne, Phys. Rev. A
\textbf{69}, 032303 (2004).

\bibitem{pepe}
P. Facchi, G. Florio, S. Pascazio, and F. Pepe, Phys. Rev. A \textbf{82}, 042313 (2010).


\end{thebibliography}
\end{document}